# Analisis dan Perancangan Wireless Roaming (Studi Kasus Universitas Baturaja)

Muhammad Sofyan[1], Leon Andretti Abdillah[2], Hadi Syahputra[3]
[1,3] Program Studi Teknik Informatika, Fakultas Ilmu Komputer, Universitas Bina Darma
[2] Program Studi Sistem Informasi, Fakultas Ilmu Komputer, Universitas Bina Darma
Palembang, Indonesia
[1]asyura38@gmail.com, [2]leon.abdillah@yahoo.com

**Abstract.** Wireless roaming is one way to improve the reliability of a network of hotspots that are still using the topology Basic Service Set (BSS). When the user walks away from one access point (AP) or one AP die then begins to lose the signal, the mobile station (MS) is automatically connected with the AP to another without reconfiguring. Devices that support wireless roaming is the AP TP-Link TL-WR740N using DD-WRT firmware that supports DHCP forwarder. Wireless roaming makes it easy for the user if there is more than one AP in an area.

## 1 Pendahuluan

Maraknya perkembangan teknologi membuat masyarakat tidak bisa lepas dari *internet*, itulah sebabnya di tempat-tempat seperti kampus atau perkantoran disediakan fasilitas *hotspot*. *Hotspot* sendiri adalah lokasi dimana *user* dapat mengakses *internet* melalui *mobile computer* (seperti *laptop* atau *smart phone*) tanpa menggunakan koneksi kabel. Jaringan *hotspot* menggunakan jaringan *wireless* yang menggunakan radio frekuensi untuk melakukan komunikasi antara perangkat komputer dengan *access point* (AP). Pada umumnya peralatan *wifi hotspot* menggunakan standardisasi IEEE 802.11b atau IEEE 802.11g dengan menggunakan beberapa tingkat keamanan seperti WEP dan atau WPA [1].

*Wireless roaming* adalah salah satu cara untuk meningkatkan reliabilitas dari suatu jaringan *hotspot* yang masih menggunakan topologi *Basic Service Set* (BSS) [2]. Ketika *user* berjalan menjauhi salah satu *access point* atau salah satu AP mati kemudian mulai kehilangan sinyal, *mobile station* (MS) secara otomatis terkoneksi dengan AP yang lain tanpa harus melakukan konfigurasi ulang. Perangkat yang mendukung *wireless roaming* adalah AP TP-Link TLWR740N dengan menggunakan *firmware* DD-WRT yang mendukung DHCP *forwarder*. *Wireless roaming* memberikan kemudahan bagi *users* jika terdapat lebih dari satu AP dalam suatu area.

Jaringan lokal nirkabel atau *wireless local area network* (*Wireless LAN* atau WLAN) merupakan jaringan komputer yang media transimisinya menggunakan gelombang radio, berbeda dengan jaringan LAN konvensional yang menggunakan kabel sebagai media transmisi sinyalnya. Standar yang digunakan dalam WLAN adalah 802.11 yang ditetapkan oleh IEEE. Versi 802.11 saat ini menyediakan kecepatan transfer data hingga 600 Mbps (802.11n). Seperti semua standar 802 IEEE, standar 802.11 berfokus pada dua lapisan terbawah model *Open System Interconnection* (OSI), yaitu physical layer dan *data link layer* [3].

Gambar 1 menunjukkan contoh penerapan jaringan WLAN. Terlihat untuk pengaksesan data tidak lagi menggunakan media kabel tetapi sudah menggunakan radio. Teknologi yang dipakai adalah *spread spectrum*. *Spread spectrum* dalam telekomunikasi adalah salah satu teknik modulasi dimana sinyal ditransimisikan dalam *bandwidth* yang jauh lebih lebar dari frekuensi sinyal awal informasi. Saat ini teknologi *spread spectrum* banyak diaplikasikan khususnya pada WLAN dan komunikasi *mobile*, karena menyediakan *bandwidth* yang lebar dan sinyalnya lebih kebal terhadap derau (*noise*) [4].

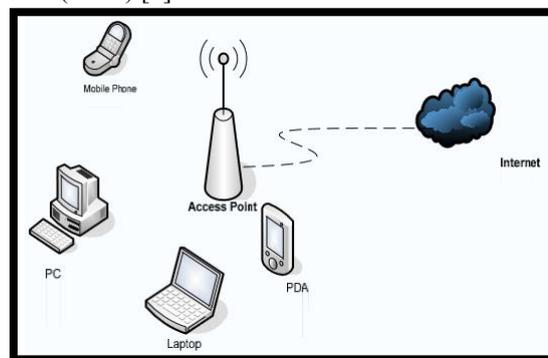

**Gambar 1.** Contoh Jaringan WLAN

Teknik nirkabel internet berbasis Wireless bertumpu pada konsep yang ditentukan oleh standart IEEE 802.11.terlepas dari jenis PHY (lapisan fisik) yang dipilih, IEEE 802.11 mendukung 3 topologi dasar untuk WLAN [5], yaitu: 1) *Independent Basic Service Set* (IBSS) atau jaringan *ad-hoc* atau *peer-to-peer*, 2) *Basic Service Set* (BSS) terdiri dari setidaknya satu AP yang terhubung ke infrastruktur jaringan kabel dan satu *set end station* nirkabel (MS). Dengan demikian, konfigurasi BSS menggunakan sebuah AP sebagai penghubung antar *client*, 3) *Extended Service Set* (ESS) terdiri dari serangkaian BSS yang saling *overlap* (masing-masing terdapat AP), yang terhubung bersama membentuk suatu *distribution system* (DS). *Mobile node* dapat melakukan *roaming* antara AP sehingga dapat mencakup kawasan yang cukup luas [6].

*Hotspot* adalah suatu koneksi jaringan *wireless* yang tersedia dan siap pakai, dimana pengguna dengan perangkat WLAN yang *compatible*, dapat terhubung ke *Internet* atau *private intranet*. *Hotspot*, atau yang lebih dikenal sebagai *Wi-Fi hotspot* tersusun atas perangkat atau komponen WLAN, *server*, dan ISP bila terhubung ke Internet [7].

## 2 Metode Penelitian

Metode yang digunakan dalam perancangan ini adalah dengan melalui tahap-tahap penelitian, seperti berikut [5]: 1) Menentukan Topologi Jaringan, 2) Menentukan Spesifikasi Perangkat, 3) Melakukan Instalasi *Software*, 4) Penentuan Lokasi Pengujian, dan 5) Melakukan Konfigurasi Jaringan.

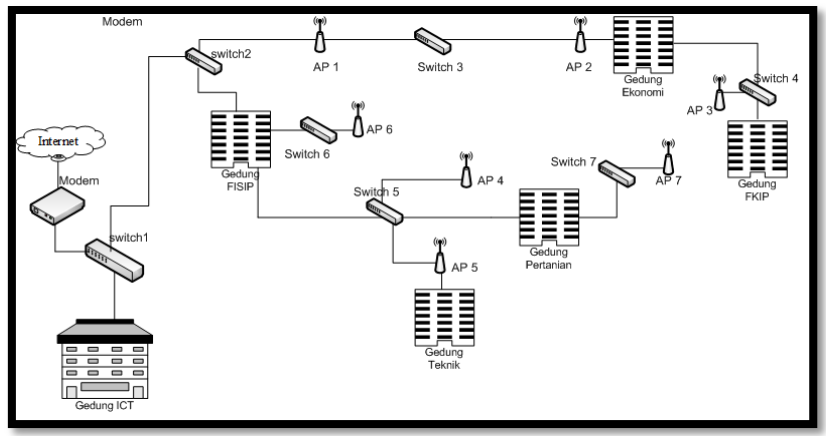

**Gambar 5.** Topologi Jaringan yang Dibangun

### 2.1 Menentukan Topologi Jaringan

Topologi jaringan yang dirancang, menggunakan *provider mobile broadband* untuk menerapkan *wireless roaming* (gambar 5). Jaringan yang dibangun akan menghubungkan 6 (enam) gedung.

### 2.2 Menentukan Perangkat *Access Point*

Untuk membangun *hotspot* yang menggunakan topologi ESS sehingga menerapkan *wireless roaming* maka digunakan dua buah *access point* yang menjalankan fungsi *DHCP forwarder*. *Access point* yang digunakan adalah TP-Link model TL-WR740N. *Access point* ini telah dilengkapi dengan perlengkapan yang dibutuhkan seperti *unit power supply* dan kabel LAN ethernet standar [5].

### 2.3 Melakukan Instalasi *Software*

Instalasi *software* dilakukan pada perangkat AP dan MS. Pada AP diinstal *firmware* DD-WRT, sedangkan pada MS diinstal Wireshark dan *bandwidth* monitor. Proses instalasi *firmware* DD-WRT pada AP dilakukan melalui dua tahapan, yaitu: 1) melakukan *upgrade* menggunakan *firmware* DD-WRT versi **factory-to-ddwrt.bin**, dan 2) **tl-wr740n- webflash.bin**.

### 2.4 Penentuan Lokasi Pemasangan *Access Point*

Dalam penelitian ini *Access Point* ditempatkan pada beberapa posisi yang berbeda dengan jarak tertentu, agar didapatkan fungsi *wireless roaming*. Perangkat ditempatkan pada beberapa gedung di Universitas Baturaja.

### 2.5 Melakukan Konfigurasi Jaringan

Setelah melakukan proses instalasi firmware DD-WRT pada AP, tahap berikutnya adalah mengkonfigurasi jaringan agar dapat beroperasi dengan baik dan dapat memenuhi syarat tercapainya jaringan *hotspot* yang menggunakan *wireless roaming*.

Konfigurasi pertama dilakukan pada *router* yang digunakan sebagai DHCP *server* sebagai berikut: 1) IP local pada router menggunakan IP address 192.168.137.1 dengan netmask 255.255.255.0 pada *interface bridge-local*, dan 2) PPP *client* yang dibuat berisi konfigurasi modem 3G yang digunakan seperti *channel modem*, APN, *username* dan *password* yang disesuaikan dengan ISP yang digunakan. Karena pada penelitian ini digunakan ISP IM3 maka APN, *username*, dan *password* diisi sesuai dengan konfigurasi dari masing-masing ISP.

Konfigurasi berikutnya dilakukan pada AP. Langkah-langkahnya adalah sebagai berikut: 1) IP local pada AP dibuat default 192.168.1.1 dengan *netmask* 255.255.255.0, 2) WAN *connection type* pada AP dibuat *disable*, 3) DHCP *type* untuk setiap *access point* dibuat menjadi DHCP *forwarder*, 4) Nama SSID untuk setiap AP dibuat sama dengan nama "Roaming", dan 5) *Security* pada setiap AP dibuat sama dengan *network key* **qwerty123**.

## 3   Hasil dan Pembahasan

Pengujian dilakukan dengan cara melakukan *test* berupa *download*, *ping time*, dan *trace route* pada MS yang terkoneksi dengan AP Roaming1 tetapi masih dalam jangkauan sinyal AP Roaming1. Setelah itu, MS dibuat menjauhi AP Roaming1 sampai diluar jangkauan AP Roaming1. Hasil dari pengujian sebelum menggunakan *wireless roaming* dapat dilihat pada Tabel 1.

**Tabel 1.** Hasil Pengujian Sebelum Menggunakan *Wireless Roaming*

| Pengujian | Dalam Jangkauan AP | Jauh dari Jangkauan AP | Diluar Jangkauan AP |
|---|---|---|---|
| Bandwidth | 360,5 Kbps | 177.8 kbps | Koneksi terputus |
| Throughput | 43.4 KB/s | 23.7 KB/s | Koneksi terputus |
| Ping Time | 160 ms | 450 ms | Koneksi terputus |

### 3.1 Pengujian *Bandwidth*, *Throughput*, dan *Ping Time* Setelah Menggunakan *Wireless Roaming*

Pengujian berikutnya mirip seperti Tabel 1, tetapi ketika MS sudah berada diluar jangkauan AP Roaming1, MS mengalami putus koneksi ke ISP tetapi tetap terhubung dengan *router*. MS secara otomatis pindah ke AP Roaming2 dan mengambil *service* dari AP Roaming2. Hasil pengujian dapat dilihat pada Tabel 2.

Dalam perpindahan tersebut terdapat delay waktu. Jika menggunakan *ping time*, terdapat beberapa kali *Request Time Out* (RTO) sedangkan jika menggunakan *download* koneksi mengalami *drop* beberapa detik sebelum kembali berjalan normal, tergantung dari *server download*.

**Tabel 2.** Hasil Pengujian Setelah Menggunakan *Wireless Roaming*

| Pengujian | Dalam Jangkauan AP | Jauh dari Jangkauan AP | Area Roaming | Diluar Jangkauan AP |
|---|---|---|---|---|
| Bandwidth | 368.5 kbps | 140.6 kbps | 0-373 kbps | 344.1 kbps |
| Throughput | 40.5 KB/s | 17.3 KB/s | 0-47.7 KB/s | 53.7 KB/s |
| Ping Time | 160 ms | 470 ms | RTO-200ms | 200ms |

Pada pengujian yang dilakukan terdapat *delay* waktu yang dibutuhkan ketika MS berpindah dari AP Roaming1 ke AP Roaming2 (0-373 kbps). Informasi tersebut mempunyai arti bandwidth yang didapat turun ke 0 kbps selama beberapa detik sebelum kembali berjalan normal. *Delay* waktu yang didapatkan ilihat pada tabel 3.

**Tabel 3.** Hasil Pengujian *Delay* Setelah Menggunakan *Wireless Roaming*

| Pengujian | Delay Waktu | RTO |
|---|---|---|
| Throughput | 43 Detik | n/a |
| Ping Time | n/a | 6 kali |

Selain pengujian menggunakan *bandwidth*, pengujian juga menggunakan *ping time* untuk mengetahui *delay* perpindahan antar AP. Ketika MS berada jauh dari jangkauan AP Roaming1 dan sudah mulai mendeteksi adanya AP lain (yang dalam pengujian ini adalah AP Roaming2, MS secara otomatis melakukan perpindahan koneksi ke AP Roaming2 yang menyebabkan beberapa kali RTO (*Request Time Out*).

### 3.2 Pengujian *Reliability* Kinerja Jaringan

Pengujian *reliability* kinerja jaringan dilakukan beberapa kali untuk memastikan apakah sistem yang dibangun sudah berjalan sesuai yang diinginkan. *Reliability* jaringan yang dimaksud adalah dimana seorang *user* yang terkoneksi dengan AP Roaming pertama tidak perlu melakukan konfigurasi ulang ketika pindah ke AP Roaming kedua dalam jaringan *wireless*. *Device* menangkap sinyal terbaik dan secara otomatis MS berpindah ke *access point* yang lain tanpa melakukan konfigurasi ulang.

### 3.3 Analisis Cara Kerja *Wireless Roaming*

*Roaming* merupakan perpindahan koneksi ketika bergerak antar *access point*. Ketika area cakupan dari dua atau lebih *access point* mengalami *overlap* maka *station* yang berada di area *overlapping* tersebut bisa menentukan koneksi terbaik, dan seterusnya mencari *access point* yang terbaik untuk melakukan koneksi. Untuk meminimalisasi *packet loss* selama perpindahan, *access point* yang lama dan *access point* yang baru saling berkomunikasi untuk mengkoordinasikan proses.

*Probe request* berisi *Service Set Identifier* (SSID) dari jaringan yang diharapkan bergabung. Ketika *access point* dengan SSID yang sama ditemukan, *access point* membalas *probe request* tersebut.

Ketika *host* menerima sebuah *beacon* yang berisi SSID dari jaringan dan berusaha bergabung. Kemudian *station* mencari alamat *MAC address* dimana *beacon* berasal mengirimkan autentifikasi *request* dengan tujuan untuk meminta *access point* agar dapat bergabung dengannya. Apabila *station* di-set untuk menerima semua macam SSID, maka *station* mencoba bergabung dengan *access point* yang pertama kali mengirimkan sinyal dan bergabung dengan *access point* yang sinyalnya paling kuat.

## 4 Kesimpulan

Berdasarkan analisis dan perancangan yang telah diuraikan di atas, penulis menyimpulkan sejumlah hal sebagai berikut:
1. Dalam hal stabilitas untuk jaringan hostpot yang menggunakan wireless roaming cukup stabil dilihat dari hasil pengujian bahwa klien dapat bergerak dan mendapatkan IP yang sama tanpa melakukan konfigurasi ulang
2. Dengan diterapkannya *wireless roaming*, jangkauan dari suatu jaringan *hotspot* dapat bertambah luas dan jumlah usernya.

## Daftar Pustaka

1. O. W. Purbo, *et al.*, "Jaringan Wireless di Dunia Berkembang," *Edisi Kedua,* 2007.
2. L. McKeag, "WLAN Roaming–the basics," *Techworld Online Magazine,* 2004.
3. M. Ergen, *Mobile broadband: including WiMAX and LTE*: Springer Science & Business Media, 2009.
4. P. Pasaribu. (2006). *Wireless LAN.* Available: http://kambing.ui.ac.id/onnopurbo/library/library-ref-ind/ref-ind-2/physical/wireless/Parlin-Publication-Wireless%20LAN-24April2006.pdf
5. F. A. K. Sejati, *et al.*, "Perancangan dan Analisis External Wireless Roaming pada Jaringan Hotspot Menggunakan Dua Jaringan Mobile Broadband," in *Seminar Nasional Teknologi Informasi dan Komunikasi Terapan (SEMANTIK2012)*, Semarang, 2012.
6. O. W. Purbo. (2001). *Gambaran Wireless LAN IEEE 802.11*. Available: http://onno.vlsm.org/v10/onno-ind-2/physical/Wireless/gambaran-wlan-ieee802-05-2001.rtf
7. D. Minoli, *Hotspot Networks: Wi-Fi for Public Access Locations*. New York: McGraw-Hill, 2003.